\documentclass[11pt,twoside]{article}
\usepackage{asp2004}
\usepackage{psfig}
\usepackage{epsf}
\usepackage{graphics}
\usepackage{lscape}
\def\edcomment#1{\iffalse\marginpar{\raggedright\sl#1\/}\else\relax\fi}
\reversemarginpar

\begin{document}
\title{Spectroscopic and Photometric Observations of SN IIn 1995N
and Energy Estimates}
\author{A. Pastorello}
\affil{INAF - Osservatorio Astrofisico di Arcetri, Largo Enrico Fermi 5, I-50125 Firenze, Italy
}
\author{I. Aretxaga}
\affil{Instituto Nacional de Astrof\'isica, \'Optica y Electr\'onica, Aptdo Postal 51 y 216, Puebla, Mexico}
\author{L. Zampieri, P. Mucciarelli, S. Benetti}
\affil{INAF - Osservatorio Astronomico di Padova, Vicolo dell' Osservatorio 5, I-25122 Padova, Italy}

\begin{abstract}
We present the results of a long--term program of monitoring of the famous SN 1995N, observed
both in photometry (\ubvrijhk ~bands) and optical spectroscopy. The observations span a period
of about 9 years. These new data, together with others available in literature, 
extend from the X--ray wavelengths to the radio, and allow to estimate the total energy radiated 
by the supernova over a decade after its explosion.
\end{abstract}
\thispagestyle{plain}

\section{Introduction}

Type IIn Supernovae (SNe IIn) form an interesting group of objects
whose observed properties (high luminosity, slow light curve decline, 
spectra with multicomponent lines, with very narrow emission features and without
broad P--Cygni absorptions) are interpreted in terms of quick reprocessing of 
the mechanical energy of the SN explosion via interaction between the SN ejecta 
and a pre--existent circumstellar medium (CSM).
Most IIn events are thought to be core--collapse explosions.
However, recently, some authors proposed that also thermonuclear
SNe may explode within a H--rich CS environment, producing IIn--like spectra 
(see SN 2002ic, e.g. Hamuy et al. 2002; Wang et al. 2004). 

SN 1995N, discovered when it was about 10 months old \citep{fox00},
 is one of the best observed SNe IIn so far. Due to its vicinity,
the peripheral location in the host galaxy and the exceptional
duration of the interaction ejecta--CSM, it is an ideal target for a long--term
monitoring at all wavelengths. SN 1995N has been studied from X--ray to radio 
wavelengths. Briefly, X--ray band observations are reported by \citet{fox00},
Mucciarelli et al. \citetext{these proceedings} and Zampieri et al. 
\citetext{in preparation}. UV--optical spectroscopy is presented by 
\citet{fran02}, late time optical photometry by \citet{li02}, and near IR 
photometry by \citet{gera02}. Moreover, a few sparse radio observations are 
available in \citet{vand96} and in the web site of K. W. Weiler and the radio--SN 
collaboration\footnote{\sl http://rsd-www.nrl.navy.mil/7213/weiler/kwdata/rsnhead.html}.

We have increased the available data set of SN 1995N with new optical and IR observations, 
obtained over a period of $\sim$9 years (1995--2004). These new data will be discussed 
in detail in a forthcoming paper \citetext{Pastorello et al., in preparation}.

\section{\ubvrijhk ~PHOTOMETRY AND OPTICAL SPECTRA}

\subsection{Optical and infrared photometry}

\begin{figure}[ht!]
\plottwo{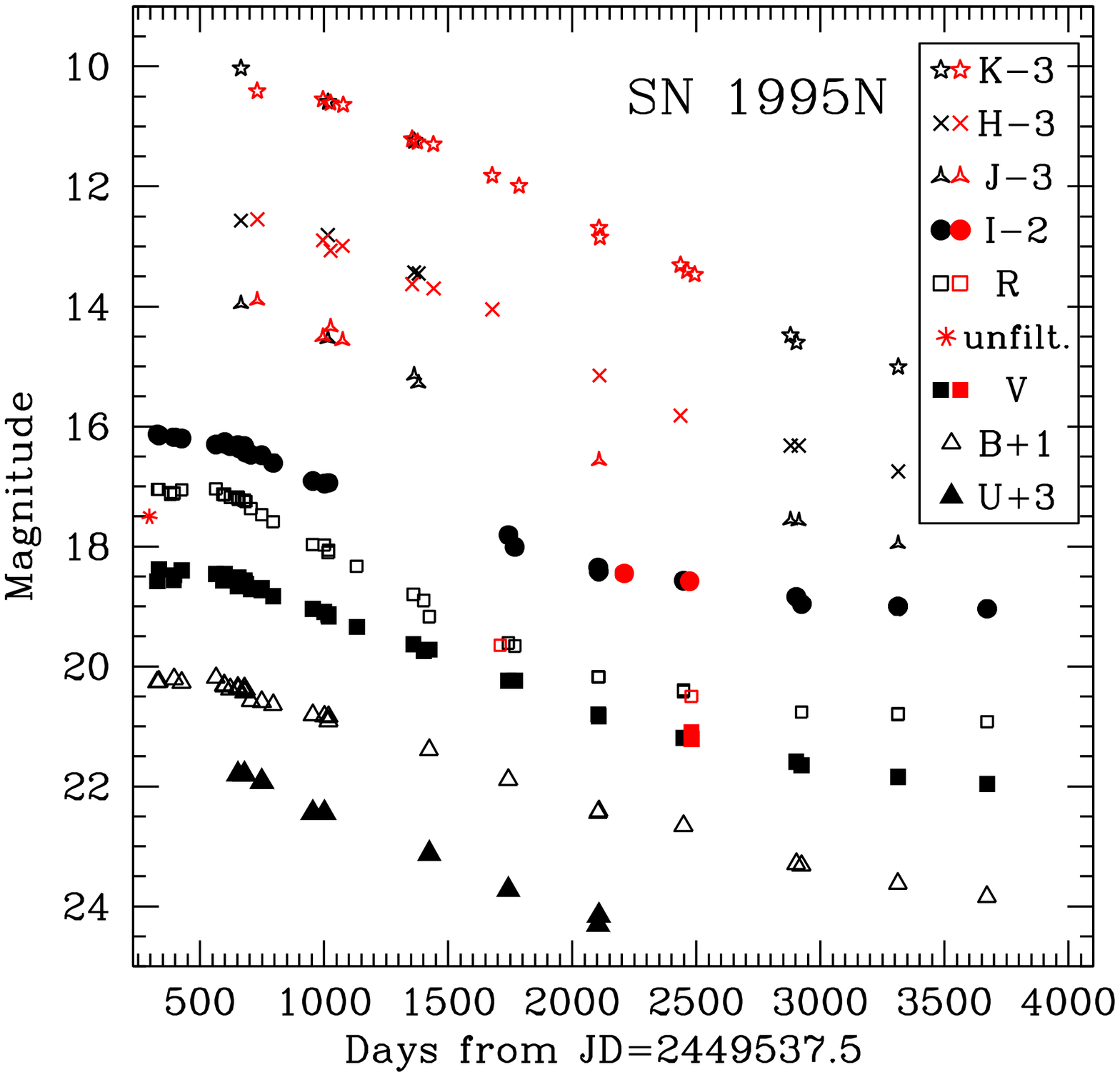}{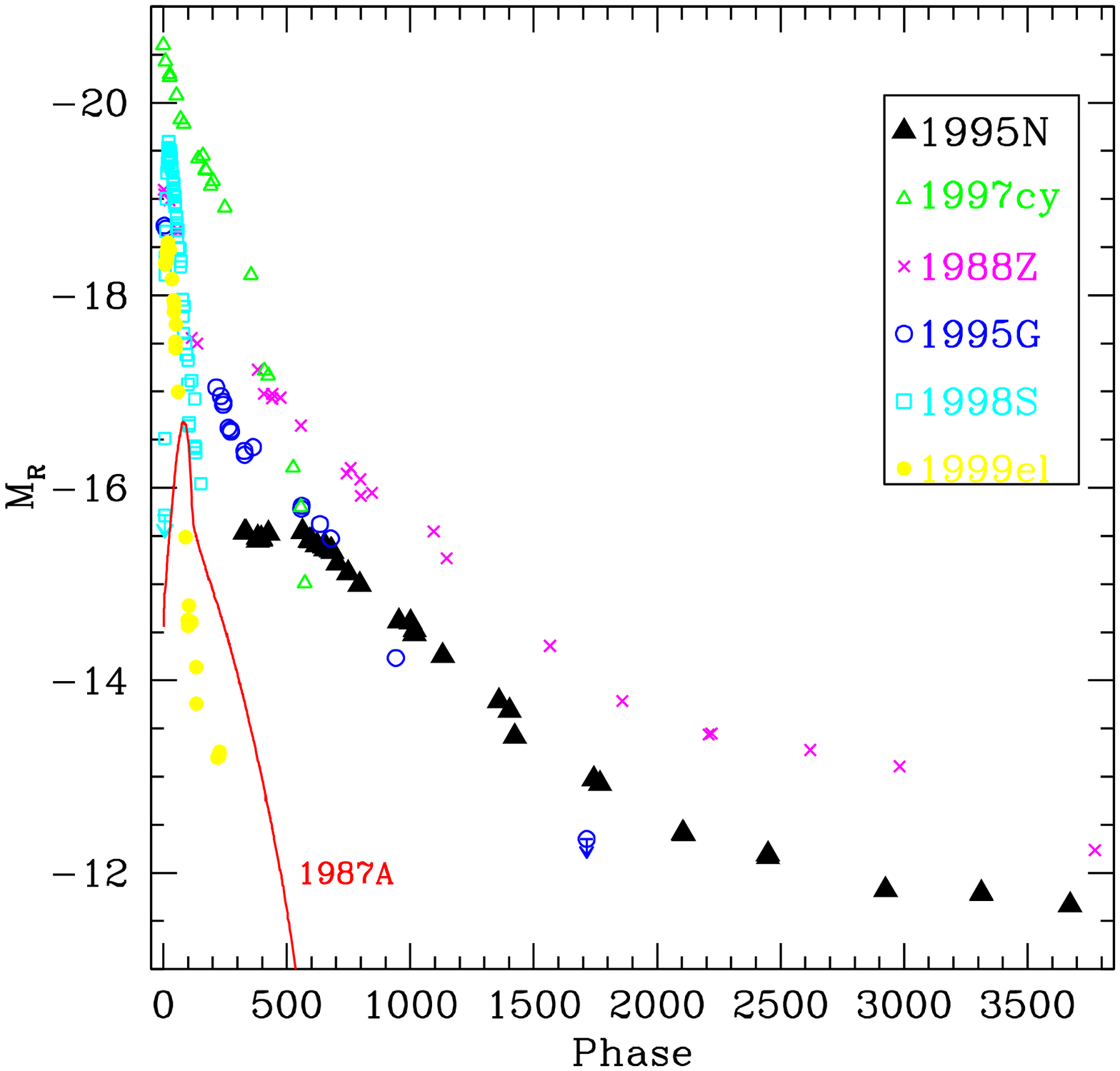}
\caption{\ubvrijhk ~light curves (left) and absolute R band light curve (right) of SN 1995N, compared
with those of other SNe IIn (see text).} \label{fig1}
\end{figure}

In Fig. \ref{fig1} (left), \ubvrijhk ~light curves of SN 1995N are reported. Together 
with our unpublished data, also data from other sources \citep{gera02,li02} 
are shown. The very slow magnitude decrease in all bands is remarkable: the observations
provide an average decline rate of about 0.10 mag/100 days (over a period of 
9 years) in V band, and $\sim$ 0.19 mag/100 days (over more than 7 years) in K band.
Another interesting property of SN~1995N and other SNe IIn, discussed also by \citet{gera02}, 
is the huge K band excess, attributed to dust emission.

Fig. \ref{fig1} (right) shows the absolute R band light curve of SN 1995N compared with those of
other interacting SNe: 1988Z (Aretxaga et al. 1999, and references therein), 1997cy \citep{germ00,tura00},
1995N \citep{pasto2}, 1998S \citep{liu00,fass00,li02} and 1999el \citep{elisa2}. 
The light curve of the non--interacting SN 1987A (Whitelock et al. 1989, and references therein) is also shown. 
Adopting the explosion epoch of SN~1995N reported by \citet{fox00}, this object is about 1 magnitude
fainter than SN~1988Z, even if their light curves are very similar in shape.\\ 
We remark that a recent observation of SN~1995N (on July 2004, i.e. $\sim$10 years after explosion), 
shows the object clearly visible in all optical bands.

\begin{figure}[ht!]
\plotone{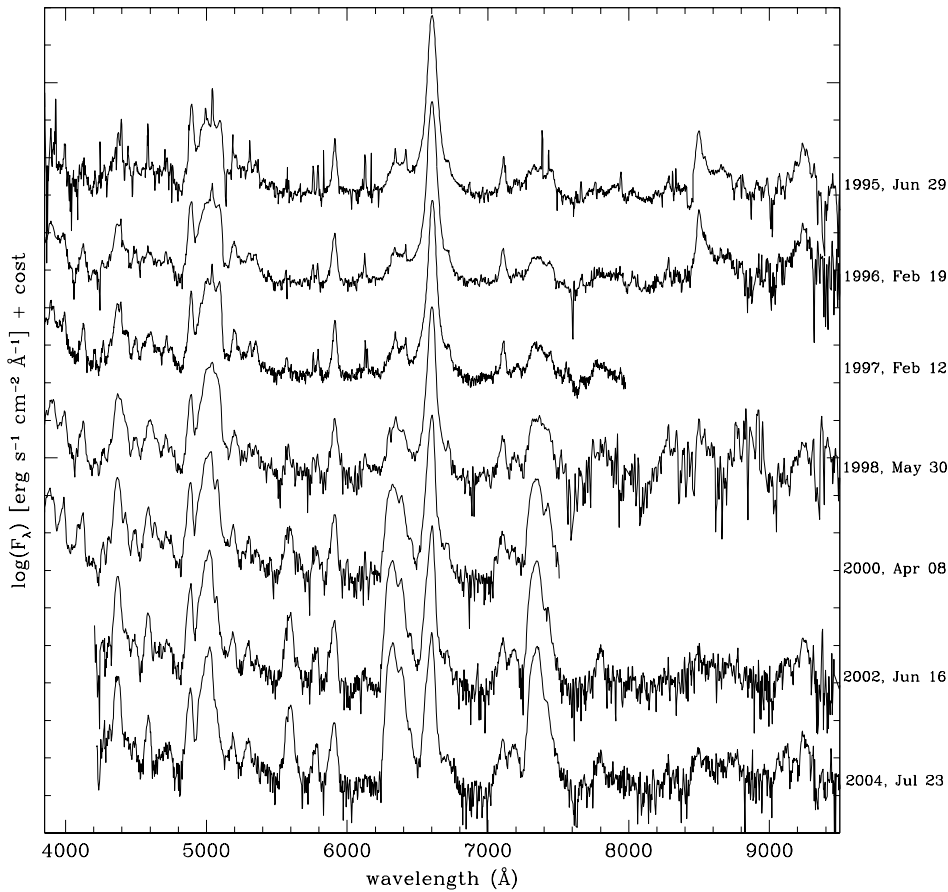}
\caption{Spectroscopic evolution of SN 1995N: in the figure only a few, representative spectra
are shown. The entire spectroscopic data set will be presented in 
Pastorello et al. (in preparation).} \label{fig2}
\end{figure}
\subsection{Optical Spectra}

A detailed analysis of the early time spectra and the line identification is 
presented by \citet{fran02}. We have obtained a large number of spectra of this object
(more than 30) during the period 1995--2004, and a small sample of our data set is shown 
in Fig. \ref{fig2}. We remark the very slow spectral evolution of this SN. The spectra are 
dominated by the H$\alpha$ emission during the first $\sim$ 4 years, then the forbidden O 
lines (especially [OIII] 4959--5007 \AA, [O I] 6300--6364 \AA, [O II] 7320--7330 \AA) increase 
in strength with time and become comparable with H$\alpha$ at about phase 8 years.
While broader components ($\ga$ 3000 km s$^{-1}$) are still visible in the last available 
spectrum, the narrow unresolved, highly ionized, forbidden lines of O, Fe, Ne, Ar, are
 detected only at early epochs, i.e. until phase of $\sim$ 3--4 years. H$\alpha$ is particularly 
prominent, and its luminosity evolution appears to be similar to that observed for 
SN~1988Z \citep{aret99}, peaking between 1 and 2 years after the SN explosion. 

\section{Energy Estimates}

After including all the data available in the literature, we estimate the X--ray to radio 
energy output of SN 1995N. We find that the major contribution to the total energy comes from 
the IR flux, as already noted for other SNe IIn \citep{gera02,pozzo04}. The K band luminosity 
is exceptionally high, especially until phase $\sim$ 2000 days. Later the X--ray luminosity 
(range 0.2--10 keV) exceeds the K band one \citetext{Zampieri et al., in preparation}.

We compute the total energy emitted in each band integrating the corresponding light curves
over the period of the available observations: the integrated energy in the U, V, I, J 
bands is about 5 $\times$ 10$^{48}$ ergs, a factor 2 times larger for the B and H bands,
4 times for the R band and a 5--6 times for the K band and the 0.2--10 keV region. 
Note also that the radio flux contributes only marginally to the total energy, lying about 
4 order of magnitudes below the optical one. Moreover we estimate the ionizing radiation 
absorbed by the cool material from the flux of H$\beta$ \citep[see][]{aret99}. 

Interpolating at missing epochs and roughly assuming that the light curve
is flat during the first unobserved 10 months, we compute the X--ray, UV ionizing 
continuum,  optical and IR energy emitted by SN~1995N over 10 years. The integration provides 
a value of E$_{TOT}$ $\sim$ 3 $\times$ 10$^{50}$ ergs. Because other SNe IIn (e.g. 1988Z) 
show a luminosity peak during the period 0--300 days, such value of E$_{TOT}$ should be 
considered as a lower limit for the total energy emitted by SN~1995N. A reasonable estimate 
is E$_{TOT} \sim$ 10$^{51}$ ergs, not far from the classical value of non--interacting 
CC--SNe and about one order of magnitude smaller than estimated for SN 1988Z \citep{aret99}.

\end{document}